\documentclass[aps,prd,onecolumn,groupedaddress,preprintnumbers,superscriptaddress,showpacs,nofootinbib]{revtex4}
\usepackage{amsmath,amssymb,latexsym}
\usepackage{graphicx}

\begin{document}

\title{
Thermodynamic and dynamical stability of Freund--Rubin compactification
}


\author{Shunichiro Kinoshita}
\email{kinoshita_at_utap.phys.s.u-tokyo.ac.jp}
\affiliation{   
Department of Physics, The University of Tokyo, Hongo 7-3-1, Bunkyo-ku,
Tokyo 113-0033, Japan
}
\author{Shinji Mukohyama}
\email{shinji.mukohyama_at_ipmu.jp}
\affiliation{
Institute for the Physics and Mathematics of the Universe (IPMU),
The University of Tokyo, 5-1-5 Kashiwanoha, Kashiwa, Chiba 277-8582, Japan
}

\preprint{UTAP-611}
\preprint{RESCEU-7/09}
\preprint{IPMU-09-0031}

\date{\today}

\begin{abstract}
 We investigate stability of two branches of Freund--Rubin
 compactification from thermodynamic and dynamical
 perspectives. Freund--Rubin compactification allows not only trivial
 solutions but also warped solutions describing warped product of 
 external de Sitter space and internal deformed sphere. We study
 dynamical stability by analyzing linear perturbations around solutions
 in each branch. Also we study thermodynamic stability based 
 on de Sitter entropy. We show complete agreement of thermodynamic and
 dynamical stabilities of this system. Finally, we interpret the results 
 in terms of effective energy density in the four-dimensional Einstein
 frame and discuss cosmological implications. 
\end{abstract}

\maketitle

 \section{Introduction and summary}
 

 De Sitter or quasi-de Sitter spacetimes describe the inflationary
 epoch of the universe at its early stages, and also the present universe
 which has entered the period of accelerated expansion.
 One of the most intriguing issues today is to realize such de Sitter vacua
 in fundamental high-energy physics.
 Particularly, higher-dimensional spacetimes are required by string theory,
 which is a promising candidate for the fundamental theory.
 In order to obtain the effective four-dimensional theory in
 higher-dimensional spacetime we should usually compactify the
 extra dimensions and stabilize the compactified internal space.
 Thus we need to embed a four-dimensional de Sitter spacetime into
 higher-dimensional spacetimes together with stabilization of the
 extra dimensions.
 
 The Freund--Rubin compactification~\cite{Freund:1980xh} is a simple
 model with a stabilization mechanism that the extra dimensions are
 dynamically compactified and stabilized by a flux of anti-symmetric
 tensor field or form field. 
 In this model with
 $(p+q)$-dimensional spacetime, a $q$-form flux field is introduced to
 stabilize the $q$-dimensional compact space. Moreover, turning on a
 positive bulk cosmological constant allows an external de Sitter space
 and an internal manifold with positive
 curvature~\cite{Bousso:2002fi,Martin:2004wp}
 (also, adding a dilaton field~\cite{Torii:2002xj}).
 Consequently, we
 obtain a $(p+q)$-dimensional product spacetime which consists of a
 $p$-dimensional de Sitter space $\mathrm{dS}_p$ and a $q$-dimensional 
 sphere $S^q$, that is the Freund--Rubin (FR) solution. 
 
 It has been known that the Freund--Rubin solution has two classes of
 dynamical instabilities~\cite{Bousso:2002fi,Martin:2004wp};
 one is attributed to homogeneous excitation ($l=0$ mode) of the
 internal space, corresponding to change of the radius of the extra
 dimensions; and the other is inhomogeneous excitation with 
 quadrupole moment ($l=2$ mode) and higher multi-pole moments ($l\ge 3$
 modes), representing deformation of the extra dimensions. 
 
 The $l=0$ mode is the so-called volume modulus or radion, and becomes 
 tachyonic when the Hubble expansion rate of the external de Sitter
 space is too large (in other words, when the flux density on the
 internal space is small). This fact implies that if the energy scale of
 the inflationary external spacetime is sufficiently larger than the 
 compactification scale of the internal space, the volume modulus will
 be destabilized. In order to avoid the emergence of this instability,
 configurations with small Hubble expansion rate is preferable.

 While instability of the volume modulus exists already in $q=2$,
 instabilities arising from deformation of extra dimensions emerge
 only if the number of extra dimensions is larger than or equal to
 four. In the unstable region, in which at least one of the $l\ge 2$
 modes are tachyonic, the external spacetime has small Hubble expansion
 rate including the Minkowski spacetime. It means that the flux
 densities are very large in the unstable region of this type. It should
 be noted that for more than four extra dimensions, the two unstable
 regions overlap so that stable configurations for the FR solution no
 longer exist.

 Little has been known about the non-perturbative properties of the
 instability from higher multi-pole modes;\footnote{The time evolution of
 unstable solutions for the $l=0$ mode was studied
 in~\cite{Krishnan:2005su}.} how it turns out after the onset of this
 instability and whether any stable configuration exists as a possible
 end-state in this model, and so on. In the previous
 work~\cite{Kinoshita:2007uk}, one of us has shown that in the
 Freund--Rubin compactification there is a new branch of solutions 
 other than the FR solutions. Those solutions are described as the
 warped product of an external de Sitter space and an internal deformed
 sphere. It has been found that the branch of the FR solutions and that
 of the warped solutions intersect at the point where the FR solution
 becomes marginally stable for the $l=2$ mode. Although we have seen
 existence of non-trivial solutions other than the FR solutions, their
 stability remains unanswered.

 In this paper we are particularly concerned with the close connection
 between dynamical stability and thermodynamic stability. The
 interesting relationship between dynamical and thermodynamic stability, 
 which is well known as the correlated stability conjecture (or the
 Gubser--Mitra conjecture~\cite{Gubser:2000mm,Gubser:2000ec}), has been
 suggested and confirmed for some 
 black objects (strings, branes and so on) by many
 authors~\cite{Reall:2001ag,Prestidge:1999uq,Gregory:2001bd,Hubeny:2002xn,Hirayama:2002hn,Miyamoto:2006nd,Miyamoto:2007mh,Brihaye:2007ju,Chen:2008vh}.
 (See e.g.~\cite{Kol:2004ww,Harmark:2007md} and references therein.)
 It is important to examine whether such connections really exist and whether
 they can be extended to systems other than black objects such as
 spacetimes with de Sitter horizons.

 In fact, for the FR solutions we can simply reinterpret the instability
 from the $l=0$ mode based on thermodynamic
 arguments~\cite{Kinoshita:2007uk} as follows (and also see 
 \cite{Kinoshita:2007ci}). For a fixed total flux on the internal space,
 the branch of FR solutions are divided into two sub-branches in terms
 of entropy defined by the total area of the de Sitter horizon: a
 sub-branch of solutions with higher entropy and the other with lower
 entropy. Both sub-branches terminate at one critical point, where the
 $l=0$ mode becomes massless and the FR solution is marginally
 stable. Moreover, the solutions on the lower-entropy sub-branch, which
 are thermodynamically unfavorable, are dynamically unstable since the
 $l=0$ mode is tachyonic. Thus thermodynamic instability exactly
 coincides with dynamical instability for the $l=0$ mode of two
 sub-branches within the FR branch.

 The aim of this paper is to examine the stability of the new branch of
 warped solutions from both dynamical and thermodynamic
 perspectives. This opens up new possibilities of the applicability of
 close connection between dynamical and thermodynamic stabilities.

 We examine the dynamical stability by analyzing perturbative stability
 of the system in a straightforward way. For simplicity, we restrict our
 considerations to the sector which behaves as scalar with respect to
 the external de Sitter space  since unstable perturbations of the FR
 solutions are in this sector.

 In the case of four-dimensional external spacetime and four-dimensional 
 internal space, we numerically obtain the Kaluza--Klein (KK) mass spectrum
 and show that the warped solutions are stable in the low Hubble
 regime, while the FR solutions are unstable due to the $l=2$ mode in
 the same regime within numerical accuracy.

 In order to reveal the thermodynamic property we derive the first law of
 de Sitter thermodynamics for Freund--Rubin compactifications. Each
 branch of the FR solutions and the warped solutions obeys the first 
 law in terms of entropy $S$ and total flux $\Phi$: 
 \begin{equation}
  \mathrm d S = - \frac{\Omega_{p-2}b}{4(p-1)h^p}\mathrm d\Phi,
 \end{equation}
 where two parameters $b$ and $h$ characterize the flux density on the
 internal space and the Hubble expansion rate of the external de Sitter
 space, respectively. This fact means that for a fixed total flux, the
 branch with higher entropy should be thermodynamically
 favored. Comparison between the entropy of the FR branch and that of
 the warped branch for a given total flux tells us which branch is
 thermodynamically favored. The result is that the warped branch is 
 entropically favored in the low Hubble regime while the FR branch is
 favored in the high Hubble regime. 
 
 The above results are briefly summarized as follows:
 \begin{itemize}
  \item For small Hubble expansion rate, the warped branch is
	thermodynamically favored and dynamically stable. 
  \item For large Hubble expansion rate, the FR branch is
	thermodynamically favored and dynamically stable. 
 \end{itemize}
 Thus, as we have expected, we see complete agreement of thermodynamic
 and dynamical stabilities for two branches of FR
 compactifications. This provides yet another example showing close
 connections between thermodynamic an dynamical properties of systems
 with horizons.

 The rest of this paper is organized as follows. 
 In Sec.~\ref{sec:Freund--Rubin_compactification} we review general
 Freund--Rubin compactifications and show the Freund--Rubin solutions
 describing $\mathrm{dS}_p\times S^q$ and the warped solutions 
 describing a warped product of an external de Sitter space and an
 internal deformed sphere. In Sec.~\ref{sec:dynamical_stability} we
 investigate dynamical stability of the warped solution by considering
 perturbations around the background solution. In
 Sec.~\ref{sec:thermodynamic_stability} we derive the first law of de
 Sitter thermodynamics and discuss thermodynamic stability for the FR
 branch and the warped branch.
 In Sec.~\ref{sec:Cosmological_implications} we interpret the above results 
 in terms of effective energy density in the four-dimensional Einstein
 frame and discuss cosmological implications.

 \section{Freund--Rubin compactification}
 \label{sec:Freund--Rubin_compactification}
 
 In this section we review general Freund--Rubin flux compactifications,
 including a bulk cosmological constant.
 We consider the $(p+q)$-dimensional action 
 \begin{equation}
  I = \frac{1}{16\pi}\int \mathrm d^{p+q}x\sqrt{-g}
   \left(R - 2\Lambda - \frac{1}{q!}F_{(q)}^2\right),
 \end{equation}
 where $\Lambda$ is a $(p+q)$-dimensional bulk cosmological constant and
 $F_{(q)}$ is a $q$-form field strength for stabilizing the
 $q$-dimensional internal manifold.
 (We use units in which $G=1$ unless otherwise noted.)
 The Einstein equation and the Maxwell equation lead to 
 \begin{equation}
  G_{MN} = \frac{1}{(q-1)!}F_{ML_1\cdots L_{q-1}}
   F_N{}^{L_1\cdots L_{q-1}}
   - \frac{1}{2q!}F_{(q)}^2 g_{MN} - \Lambda g_{MN}
 \end{equation}
 and
 \begin{equation}
  \nabla_M F^{MN_1\cdots N_{q-1}} = 0,
 \end{equation}
 where the $q$-form field satisfies the Bianchi identity 
 $\nabla_{[M_1} F_{M_2\cdots M_{q+1}]} = 0$.

  \subsection{Freund--Rubin solution}
  
  These equations have well-known solutions originally found by Freund
  and Rubin~\cite{Freund:1980xh}. The metric and the $q$-form field
  strength in the Freund--Rubin solutions are given by 
  \begin{equation}
   \mathrm ds^2 = - \mathrm dt^2 + e^{2ht}\mathrm d\vec{x}_{p-1}^2 
    + \rho^2 \mathrm d\Omega_{q}^2,
  \end{equation}
  and 
  \begin{equation}
   F_{(q)} = b\epsilon_{\mu_1 \cdots \mu_q},
  \end{equation}
  where $\epsilon_{\mu_1 \cdots \mu_q}$ is the volume element of the
  $q$-sphere with a radius $\rho$.
  These solutions describe the direct product of a $p$-dimensional external
  de Sitter space with Hubble expansion rate $h$ and a $q$-sphere with
  radius $\rho$. The internal $q$-sphere is supported by the $q$-form
  flux with a flux density $b$. The Einstein equation and Maxwell
  equation yield relations among $b$, $h$ and $\rho$: 
  \begin{equation}
   (p-1)(p+q-2)h^2 + (q-1)b^2 = 2\Lambda,
  \end{equation}
  \begin{equation}
   (q-1)^2 \rho^{-2} + (p-1)^2 h^2 = 2\Lambda.
  \end{equation}
  Evidently, these relations allow a one-parameter family of
  solutions.

  \subsection{Warped solution}

  There is another one parameter family of non-trivial solutions. Their
  geometry is a warped product of de Sitter space and a deformed
  sphere. Such warped solutions are described by the following ansatz
  for the metric
  \begin{equation}
   \mathrm ds^2 = e^{2\phi(r)}[- \mathrm dt^2 + e^{2ht} 
    \mathrm d\vec x_{p-1}^2]
    + e^{-\frac{2p}{q-2}\phi(r)}
    [\mathrm dr^2 + a^2(r) \mathrm d\Omega^2_{q-1}],
    \label{eq:metric_ansatz}
  \end{equation}
  and the $q$-form flux 
  \begin{equation}
   F_{(q)} = b e^{-\frac{2p(q-1)}{q-2}\phi}a^{q-1} \mathrm dr \wedge
    \mathrm d\Omega_{q-1},\label{eq:q_form}
  \end{equation}
  where $b$ is a constant.
  Note that (\ref{eq:q_form}) automatically satisfies the Maxwell
  equation and the Bianchi identity. Then, from the Einstein equation we
  have the two equations,
  \begin{equation}
   \begin{aligned}
    \frac{a''}{a} &= \frac{b^2}{q - 2}
    e^{- \frac{2p(q - 1)}{q - 2}\phi}
    + 
    \frac{p(p - 1)}{q - 2}
    h^2 e^{-\frac{2(p + q - 2)}{q - 2}\phi}
    - 
    \frac{2\Lambda}{q - 2} e^{- \frac{2p}{q - 2}\phi}
    -
    (q - 2)\frac{{a'}^2 - 1}{a^2},\\
    \phi'' &= (p - 1)h^2
    e^{-\frac{2(p + q - 2)}{q - 2}\phi}
    + \frac{q - 1}{p + q - 2}
    b^2 e^{- \frac{2p(q - 1)}{q - 2}\phi}
    - 
    \frac{2\Lambda}{p + q - 2}e^{- \frac{2p}{q - 2}\phi} - 
    (q - 1)\frac{a'}{a}\phi' ,
   \end{aligned}
  \end{equation}
  and the constraint equation
  \begin{equation}
   \frac{(q-1)(q-2)}{2}\left[\left(\frac{a'}{a}\right)^2 -
			\frac{1}{a^2}\right] 
   = \frac{p(p+q-2)}{2(q-2)} \phi'^2
   + \frac{p(p-1)}{2}h^2 e^{-\frac{2(p+q-2)}{q-2}\phi}
   + \frac{b^2}{2}e^{-\frac{2p(q-1)}{q-2}\phi}
   - \Lambda e^{-\frac{2p}{q-2}\phi}.
  \end{equation} 
  We are interested in the case where the internal space is
  compact. Thus, we consider an interval $r_-\leq r\leq r_+$, where
  $a(r)$ vanishes at the endpoints, $a(r_{\pm})=0$, and is positive
  between them. Then the regularity requires the following boundary
  conditions: 
  \begin{equation}
  |a'(r_\pm)| = 1, \quad \phi'(r_\pm) = 0
  \end{equation}
  at the endpoints. These conditions ensure that the internal space with
  spherical topology is regular at the north and south poles.

\begin{figure}[t]
   \begin{center}
    \includegraphics[width=.48\linewidth]{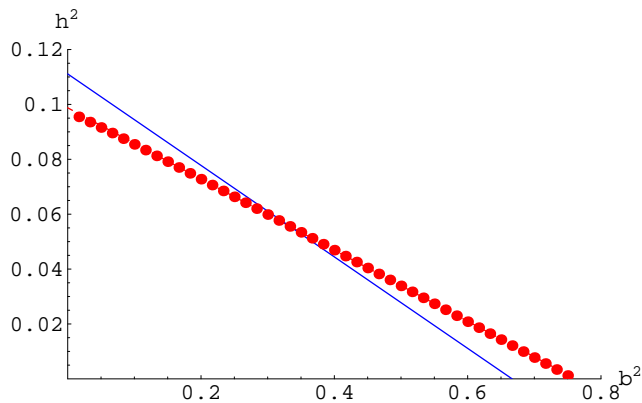}
    \caption{Two branches of solutions in the $(b^2, h^2)$ plane. The
    blue solid line represents the branch of Freund--Rubin
    solutions. The red bold points correspond to values calculated
    numerically for which warped solutions have been found. Two branches
    intersect at one point $(b^2, h^2) = (\frac{1}{3},\frac{1}{18})$. 
    Note that we have set $\Lambda = 1$.}
    \label{fig:b2h2}
   \end{center}
  \end{figure}

  A one-parameter family of solutions to the above equations and
  boundary conditions for $p=4$ and $q=4$ was found
  numerically by one of the authors~\cite{Kinoshita:2007uk}. This warped
  branch of solutions emanates from the marginally stable solution in
  the branch of Freund--Rubin solutions as shown in
  Fig~\ref{fig:b2h2}. In Fig~\ref{fig:b2h2}, two lines representing two
  branches intersect at one point. At the intersection, the solution in
  the warped branch is no more warped and reduces to a FR solution. For
  $b^2$ smaller than the value at the intersection, the internal space
  is prolate. On the other hand, for $b^2$ larger than
  the value at the intersection, the internal space is oblate. 
  Therefore, while the numerically obtained value of $b^2$ at
  the intersection includes numerical errors, the statement that the
  warped solution reduces to the FR solution at the intersection is
  exact: there must be a boundary between the oblate and prolate cases;
  and at the boundary, the internal space is spherical and the warp
  factor is trivial. 

  In the following sections, we shall investigate stability of the FR
  and warped branches.

 \section{Dynamical stability}
 \label{sec:dynamical_stability}
 
 In this section we investigate dynamical stability of FR solutions and
 warped solutions by considering linear perturbations around them. We 
 will concentrate on scalar-type perturbation with respect to the
 external de Sitter space since in the case of the FR solution
 instability arises from perturbations of this type. 
 
  \subsection{Background}

  In the previous section we have already shown the background metric
  and form field. In this subsection we rewrite them in a form which is
  convenient for the analysis of perturbation equations. 
  
  We suppose that the $(p+q)$-dimensional metric is given by 
  \begin{equation}
   \mathrm ds^2 = A^2(y)g_{\mu\nu}(x)\mathrm dx^\mu\mathrm dx^\nu
    + \mathrm dy^2
    + B^2(y)\gamma_{ij}(z)\mathrm dz^i\mathrm dz^j,
  \end{equation}
  where $g_{\mu\nu}$ is the metric of $p$-dimensional Lorentzian
  Einstein space and $\gamma_{ij}$ is the metric of $(q-1)$-dimensional
  Euclidean Einstein space.
  Note that the Riemann tensors with respect to $g_{\mu\nu}$ and
  $\gamma_{ij}$ respectively satisfy 
  ${}^{(g)}R_{\mu\nu} = K(p-1)g_{\mu\nu}$ and 
  ${}^{(\gamma)}R_{ij} = k(q-2)\gamma_{ij}$.
  The $q$-form field strength is 
  \begin{equation}
   F_{(q)} = \frac{b}{A^p}B^{q-1}\sqrt{\gamma}\mathrm dy
    \bigwedge_{i=1}^{q-1}\mathrm dz^i,
  \end{equation}
  which satisfies the Maxwell equation and the Bianchi identity
  automatically.

  Then, non-vanishing components of the $(p+q)$-dimensional Einstein
  tensor $G_{MN}$ become 
  \begin{equation}
   \begin{aligned}
    G_{\mu\nu} &= 
    \left[
    \frac{(p-1)(p-2)}{2}\frac{{A'}^2 - K}{A^2} + (p-1)\frac{A''}{A}
    + \frac{(q-1)(q-2)}{2}\frac{{B'}^2 - k}{B^2} + (q-1)\frac{B''}{B}
    + (p-1)(q-1)\frac{A'B'}{AB}
    \right]A^2g_{\mu\nu},\\
    G_{yy} &=
    \frac{p(p-1)}{2}\frac{{A'}^2 - K}{A^2}
    + \frac{(q-1)(q-2)}{2}\frac{{B'}^2 - k}{B^2}
    + p(q-1)\frac{A'B'}{AB},\\
    G_{ij} &=
    \left[
    \frac{p(p-1)}{2}\frac{{A'}^2 - K}{A^2} + p\frac{A''}{A}
    + \frac{(q-2)(q-3)}{2}\frac{{B'}^2 - k}{B^2} + (q-2)\frac{B''}{B}
    + p(q-2)\frac{A'B'}{AB}
    \right]B^2\gamma_{ij},
   \end{aligned}
  \end{equation}
  (here in this section the prime denotes the derivative with
  respect to $y$)
  and the energy-momentum tensor of the $q$-form flux field is given by 
  \begin{equation}
   T_{\mu\nu} = - \frac{b^2}{2A^{2(p-1)}}g_{\mu\nu}, \quad
    T_{yy} = \frac{b^2}{2A^{2p}}, \quad
    T_{ij} = \frac{b^2}{2A^{2p}}B^2\gamma_{ij},
  \end{equation}
  and the other components vanish.
  
  Finally, we note that if we set $A = e^\phi$, 
  $B = e^{-\frac{p}{q-2}\phi}a$, 
  $\mathrm dy = e^{-\frac{p}{q-2}\phi}\mathrm dr$, $k=1$ and $K=h^2$ in
  the above equations, then all equations for the background ansatz in
  Sec.~\ref{sec:Freund--Rubin_compactification} are reproduced. 
  
  \subsection{Perturbation}

  We now leave the subject of background and turn our attention to
  linear perturbations around the background. We suppose that the 
  $p$-dimensional external spacetime and the $(q-1)$-dimensional
  internal space are the Einstein manifolds with the metric $g_{\mu\nu}$
  and $\gamma_{ij}$, respectively. In this case we can decompose tensors
  on the $(p+q)$-dimensional spacetime into scalar-type, vector-type and
  tensor-type components with respect to $g_{\mu\nu}$ and
  $\gamma_{ij}$. Hence, in our analysis we decompose perturbations of
  the metric and the form field into different types and obtain
  decoupled perturbation equations in each sector. 
  
  As we already mentioned before, we will concentrate on scalar-type
  perturbations. Especially, we suppose that the internal manifold is
  topologically a sphere and has $\mathrm{SO}(q)$-isometry, namely 
  $\gamma_{ij}(z)\mathrm dz^i\mathrm dz^j = \mathrm d\Omega_{q-1}^2$
  which is the metric of unit round $(q-1)$-sphere.
  Then what we are interested in is perturbations which are scalar-type
  quantities with respect to not only $p$-dimensional de Sitter symmetry
  of the background external space but also $\mathrm{SO}(q)$ symmetry of the
  background internal space.
  
  In this paper, for simplicity we assume that the perturbations
  preserve the $\mathrm{SO}(q)$ symmetry of the background internal space. 
  By choosing an appropriate gauge (see Appendix~\ref{app:gauge}), 
  we can write the perturbed metric and field strength as follows:
  \begin{equation}
   \mathrm ds^2 = \left(1+\Pi\mathsf Y\right)A^2(y)g_{\mu\nu}
    \mathrm dx^\mu\mathrm dx^\nu
    + \left[1+(\Pi-\Omega)\mathsf Y\right]\mathrm dy^2
    + \left[1+\left(\frac{\Omega}{q-1}-\frac{p-1}{q-1}\Pi\right)
       \mathsf Y\right]
    B^2(y)\mathrm d\Omega_{q-1}^2,
  \end{equation}
  and
  \begin{equation}
   F_{(q)} = b \frac{B^{q-1}}{A^p}
    \mathrm dy\wedge\mathrm d\Omega_{q-1} 
    + \mathrm d\left(\varphi\mathsf Y\right)\wedge\mathrm d\Omega_{q-1}
  \end{equation}
  where $\mathsf Y(x)$ is scalar harmonics on the de Sitter space with
  the Hubble expansion rate $h$ and variables $\Pi$, $\Omega$ and
  $\varphi$ depend on only $y$-coordinate due to the
  $\mathrm{SO}(q)$-symmetry.

  From the linearized Einstein equation and Maxwell equation, we obtain
  a set of two perturbation equations for two variables $\Pi$ and $\Omega$: 
  \begin{equation}
   \begin{aligned}
    (p+q-2)\Pi'' + (q-2)\Omega'' 
    +& (p+q-2)\left[p\frac{A'}{A} - (q-1)\frac{B'}{B}\right]\Pi'
    + (q-2)\left[p\frac{A'}{A} + (q-1)\frac{B'}{B}\right]\Omega'\\
    +& \left[\frac{\mu^2}{A^2} + \frac{2(q-2)}{B^2}\right]
    \left[(p+q-2)\Pi - q\Omega\right] = 0,\\
    \Omega''
    + \left[(3p-2)\frac{A'}{A} + 3(q-1)\frac{B'}{B}\right]\Omega'
    -& \left[\frac{2(p+q-2)(q-2)}{B^2} - 4\Lambda\right]\Pi\\
    +& \left[\frac{\mu^2 + 2h^2(p-1)^2}{A^2}
    + \frac{2q(q-2)}{B^2} - 4\Lambda\right]\Omega = 0,
   \end{aligned}\label{eq:perturbation_eq}
  \end{equation}
  where $\mu^2$ is the KK mass squared which is defined by 
  $\nabla^2 \mathsf{Y}(x) = \mu^2 \mathsf{Y}(x)$.
  Here $\nabla_\mu$ denotes the covariant derivative with respect to
  $g_{\mu\nu}$.
  Moreover we have an algebraic equation for $\varphi$:
  \begin{equation}
   \varphi =  \frac{A^pB^{q-1}}{2b}
    \left[(p+q-2)\frac{B'}{B}\Pi
    - \frac{1}{A^{p-2}B^q}\left(A^{p-2}B^q\Omega\right)'\right].
  \end{equation}
  Boundary conditions are specified by the regularity at the poles of the
  internal space of spherical topology, which are characterized by
  $B=0$. They are given by 
  \begin{equation}
   (p+q-2)\Pi - q\Omega = \Omega' = 0 \quad \mbox{at} \quad B=0.
    \label{eqn:boundary-cond}
  \end{equation}
  Thus the perturbation equations are reduced to eigenvalue problems
  with eigenvalue $\mu^2$. 
  If the spectrum of $\mu^2$ is non-negative, we can conclude that the
  background spacetime is dynamically stable.

  Before we discuss stability of the warped solution, let us recall
  stability of the FR solution.
  For the FR solution, we set $A=1$ and $B=\rho \sin \frac{y}{\rho}$.
  Then the perturbation equations reduce to 
  \begin{equation}
   \begin{aligned}
    (p+q-2)\Pi'' + (q-2)\Omega'' 
    -& (p+q-2)(q-1)\frac{B'}{B}\Pi'
    + (q-2)(q-1)\frac{B'}{B}\Omega'\\
    +& \left[\mu^2 + \frac{2(q-2)}{B^2}\right]
    \left[(p+q-2)\Pi - q\Omega\right] = 0,\\
    \Omega''
    + 3(q-1)\frac{B'}{B}\Omega'
    -& \left[\frac{2(p+q-2)(q-2)}{B^2} - 4\Lambda\right]\Pi\\
    +& \left[\mu^2 + 2h^2(p-1)^2
    + \frac{2q(q-2)}{B^2} - 4\Lambda\right]\Omega = 0.
   \end{aligned}
  \end{equation}
  Eliminating $\Omega$ from these equations we obtain a single forth
  order differential equation for $\Pi$, 
  \begin{equation}
   D^2\cdot D^2 \Pi + 
    2\left[\mu^2 + (p-1)h^2 + \frac{p(q-1)}{p+q-2}b^2\right]D^2 \Pi
    + \mu^2\left[\mu^2 + 2(p-1)h^2 - \frac{2(q-1)(p-2)}{p+q-2}b^2
	   \right]\Pi= 0,
  \end{equation}
  where $D^2$ denotes Laplacian on $S^q$ with a radius $\rho$.
  We expand $\Pi$ in terms of scalar hamonics $Y(y)$ on $S^q$, and then
  the mass eigenvalues $\mu^2$ are given by 
  \begin{equation}
   \begin{split}
    \mu^2_\pm =& \lambda + \frac{(q-1)(p-2)}{p+q-2}b^2 - (p-1)h^2\\
     &\pm \sqrt{\left[\frac{(q-1)(p-2)}{p+q-2}b^2- (p-1)h^2\right]^2
    + \frac{4(q-1)(p-1)}{p+q-2}b^2\lambda},
   \end{split}\label{eq:mass_spectrum}
  \end{equation}
  where $D^2 Y(y) = -\lambda Y(y)$ with 
  $\lambda = l(l+q-1)\rho^{-2}$~\cite{Bousso:2002fi,Martin:2004wp}.
  It is clear from this expression that the scalar perturbations for
  each multi-pole moment $l$ generically have two independent modes
  corresponding to the mass eigenvalues $\mu_+^2$ and $\mu_-^2$. 
  However, for $l=0$ and $l=1$, only one of them is physical and the
  other is a gauge mode. 
  For $l \ge 2$, we denote the modes with the mass squared $\mu_+^2$ and 
  $\mu_-^2$ as $l=2_+,3_+,\cdots$ and $l=2_-,3_-,\cdots$, respectively.
  In the $l=1$ case the physical mode has the mass eigenvalue
  $\mu_+^2(l=1)$. On the other hand, in the $l=0$ case we have a
  physical mode with 
  $\mu^2(l=0) = 2\frac{(q-1)(p-2)}{p+q-2}b^2 - 2(p-1)h^2$, 
  corresponding to $\mu_+^2$ for 
  $\frac{(q-1)(p-2)}{p+q-2}b^2 > (p-1)h^2$ and $\mu_-^2$ for
  $\frac{(q-1)(p-2)}{p+q-2}b^2 < (p-1)h^2$.   

\begin{figure}[t]
   \begin{center}
    \includegraphics[width=.48\linewidth]{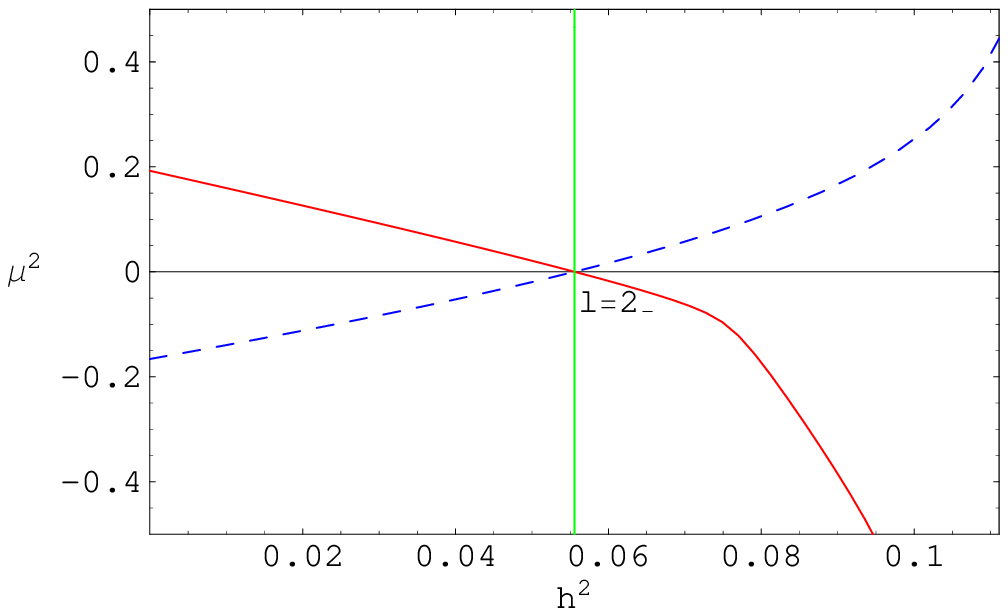}
    \includegraphics[width=.48\linewidth]{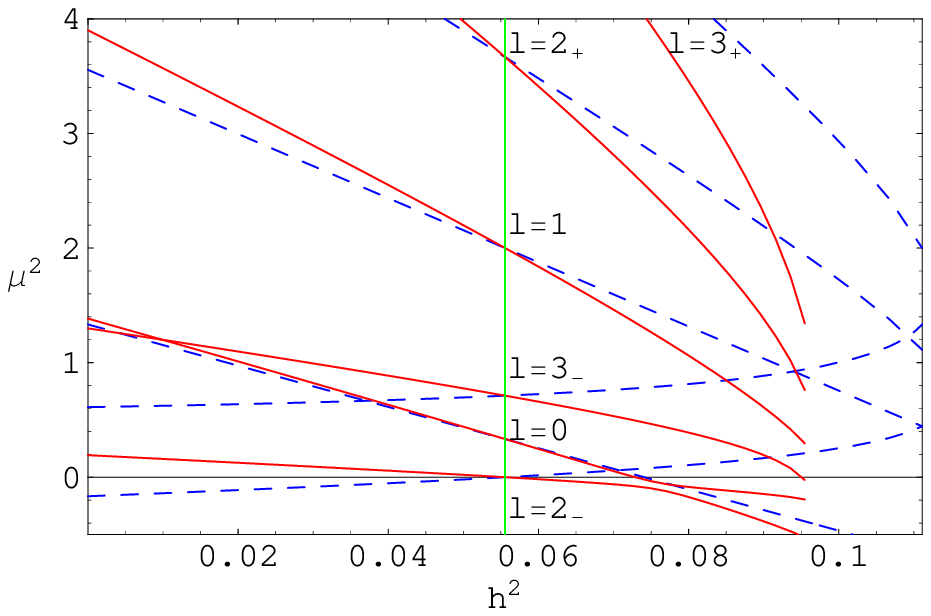}
    \caption{The mass spectrum for scalar perturbations: the $l=2_-$
    mode (left) and some modes with $l=0,1,2_\pm,3_\pm$ (right). 
    Red solid lines
    and blue dashed lines indicate the warped branch and the FR branch,
    respectively. The green vertical line indicates the critical value
    $h^2 = h^2_{\mathrm{c} (l=2)}=\Lambda/18$ at which two branches
    merge. In the low Hubble region where the FR branch is unstable, the
    warped branch is stable. }
    \label{fig:KK_mass}
   \end{center}
  \end{figure}

  This mass spectrum leads to the following result for dynamical
  stability of the FR solutions; 
  when $h^2>h_{\mathrm{c}(l=0)}^2$, where 
  \begin{equation}
   h_{\mathrm{c}(l=0)}^2 = \frac{2\Lambda(p-2)}{(p-1)^2(p+q-2)},
    \label{eqn:hcrit0-dynam}
  \end{equation}
  the $l=0$ mode is tachyonic and the FR solution is unstable arising
  from homogeneous excitation of the sphere. 
  In addition, for $q \ge 4$, when $h^2$ becomes smaller than the
  critical value $h_{\mathrm{c}(l=2)}^2$ given by 
  \begin{equation}
   h^2_{\mathrm{c} (l=2)} = 
    \frac{2\Lambda[(p-1)q^2 - (3p-1)q + 2]}{q(q-3)(p-1)^2(p+q-2)},
    \label{eqn:hcrit2-dynam}
  \end{equation}
  the mass squared $\mu_-^2$ is negative for $l=2$ and the FR solution
  is unstable arising from inhomogeneous excitations. As we have
  explained above, we call this mode $l=2_-$. Some modes with $l\geq 3$
  can be unstable when $h^2$ is even smaller. 
  The mass spectrum for the scalar perturbations of the Freund--Rubin
  solution is shown as blue dashed lines in 
  Fig~\ref{fig:KK_mass}. 

  Now, let us consider the warped solutions.
  Especially we would like to examine dynamical stability of the warped
  solutions in the small Hubble regime in which the FR solutions suffer
  from instability of the inhomogeneous excitations. 
  For the warped solution we numerically solve the eigenvalue
  problem for a set of differential equations (\ref{eq:perturbation_eq})
  with the boundary condition (\ref{eqn:boundary-cond}) in order to find
  the mass spectrum $\mu^2$. 
  In Fig~\ref{fig:KK_mass}, we present the numerically obtained mass 
  spectrum of the warped branch for $p=4$ and $q=4$.
  The red solid lines represent $\mu^2$ for the warped branch and the
  blue dashed lines for the FR branch.
  In the left panel of Fig~\ref{fig:KK_mass}, we focus our attention 
  to the $l=2_-$ mode since it is the first unstable mode for
  inhomogeneous perturbations on the FR branch. 
  Note that ``$l=2$'' means quadrupole moment with respect to the 
  $\mathrm{SO}(q+1)$-symmetry of the internal space and that, rigorously
  speaking, this terminology is valid only for the FR branch. 
  However, as explained in the end of
  Sec.~\ref{sec:Freund--Rubin_compactification}, there is a critical
  value of $b^2$ at which the solution in the warped branch reduces to
  an unwarped, FR solution. Since $h^2$ is determined by $b^2$ in 
  each branch, this implies that there is a critical value of $h^2$ at
  which the solution in the warped branch and that in the FR branch
  represent the same solution. Therefore, it makes perfect sense to
  define the ``$l=2$ mode'' for the warped branch as the mode which
  approaches the $l=2$ mode of the FR branch as $(b^2,h^2)$ approaches
  the critical value. We can define $l=3_{\pm}, 4_{\pm}, \cdots$ modes
  for the warped branch in a similar way. As one can see in the left
  panel of Fig~\ref{fig:KK_mass},  for 
  $h^2<h^2_{\mathrm{c}(l=2)} = \Lambda/18$, 
  $\mu^2$ for the $l=2_-$ mode of the warped branch is positive, while
  the $l=2_-$ mode of the FR branch becomes tachyonic. Thus the warped 
  branch is stable configuration in the low Hubble region, in which the
  FR branch is unstable. Actually, within numerical accuracy one can see
  that the red solid line and the blue dashed line for $l=2_-$ intersect
  at $\mu^2=0$. This implies that the critical value of $h^2$ at which
  the warped solution reduces to the FR solution agrees with
  (\ref{eqn:hcrit2-dynam}).

  The mass squared for some other modes as well as the $l=2_-$ mode is
  shown in the right panel of Fig~\ref{fig:KK_mass}. As seen from the
  mass spectrum, the warped branch has no unstable mode in the low
  Hubble region. In addition we notice that $\mu^2$ of the warped branch
  (the red solid line) is larger than that of the FR branch (the blue
  dashed line) for $h^2$ smaller than $h^2_{\mathrm{c} (l=2)}$ (the left
  hand side). This means that deformation of the internal space and
  warping tend to stabilize the shape modulus of the internal space in
  the low Hubble region. In other words, the tachyonic shape modulus is
  stabilized by the condensation of the modulus itself.

 \section{Thermodynamic stability}
 \label{sec:thermodynamic_stability}
 
 In the previous section we have investigated dynamical stability of
 two branches of Freund--Rubin compactification. In this section we
 shall investigate thermodynamic stability of the same system and
 compare the results with dynamical stability.

  \subsection{Thermodynamic relations}
  
  In Freund--Rubin compactifications we can define various physical
  quantities characterizing thermodynamic properties of the system.  
  One of the most important among them is the de Sitter entropy $S$,
  which is defined by one quarter of the total area $\mathcal A$ of de
  Sitter horizon. For the metric (\ref{eq:metric_ansatz}) it is given by 
  \begin{equation}
   S \equiv \frac{\mathcal A}{4}
    = \frac{\Omega_{p-2}\Omega_{q-1}}{4h^{p-2}}
    \int^{r_+}_{r_-} \mathrm dr \,e^{-\frac{2(p+q-2)}{q-2}\phi} a^{q-1}.
    \label{eq:entropy}
  \end{equation}
  Also, we can define the total flux of the $q$-form field
  (\ref{eq:q_form}) as 
  \begin{equation}
   \Phi 
    \equiv \oint F_{(q)}
    = b \Omega_{q-1}\int^{r_+}_{r_-} \mathrm dr \,
    e^{-\frac{2p(q-1)}{q-2}\phi} a^{q-1},
  \end{equation}
  which is a conserved quantity for this system.
  

Before examining thermodynamic stability, let us derive the first law of
de Sitter thermodynamics for Freund--Rubin compactifications.

For this purpose it is convenient to consider variations of the
Euclidean action for the system since the on-shell Euclidean action is
directly related to the de Sitter entropy as we shall see below. 
  Assuming $\mathrm{SO}(p+1) \times \mathrm{SO}(q)$ isometry, we can
  take the metric ansatz as 
  \begin{equation}
   \mathrm ds^2_\mathrm{Euclid} = e^{2\phi(r)}h^{-2}\mathrm d\Omega^2_p
    + e^{-\frac{2p}{q-2}\phi(r)}
    [\mathrm dr^2 + a^2(r) \mathrm d\Omega^2_{q-1}],
  \end{equation}
  where $\mathrm d\Omega^2_p$ and $\mathrm d\Omega^2_{q-1}$ denote the
  metrics of the unit round $p$- and $(q-1)$-sphere, respectively.
  The $q$-form field strength is given by  
  \begin{equation}
   F_{(q)} = \psi'(r)\mathrm dr\wedge\mathrm d\Omega_{q-1}.
  \end{equation}
  The Euclidean action is given by 
  \begin{equation}
   I_\mathrm{Euclid} = - \frac{1}{16\pi}\int\mathrm d^{p+q}x_\mathrm{E} 
    \sqrt{g_{\mathrm E}}
    \left(R - 2\Lambda - \frac{1}{q!}F_{(q)}^2\right).
  \end{equation}
  The Ricci scalar and the field strength are 
  \begin{equation}
   \begin{aligned}
    R =& e^{\frac{2p}{q-2}\phi}
    \left[
    (q-1)(q-2)\frac{{a'}^2 + 1}{a^2}
    - \frac{p(p+q-2)}{q-2}{\phi'}^2 + p(p-1)h^2e^{-\frac{2(p+q-2)}{q-2}\phi}
    \right]\\
    & - \frac{2e^{\frac{2p}{q-2}\phi}}{a^{q-1}}
    \left\{
    a^{q-1}\left[(q-1)\frac{a'}{a} - \frac{p}{q-2}\phi'\right]
    \right\}',
   \end{aligned}
  \end{equation}
  and 
  \begin{equation}
   - \frac{1}{q!}F_{(q)}^2
    = - \frac{e^{\frac{2pq}{q-2}\phi}}{a^{2(q-1)}}{\psi'}^2.
  \end{equation}
  Hence we have 
the following expression for 
the Euclidean action 
  \begin{equation}
   \begin{aligned}
    I_\mathrm{Euclid}[a,\phi,\psi]
    =& - \frac{\Omega_p\Omega_{q-1}}{16\pi h^p}
    \int^{r_+}_{r_-}\mathrm dr
    \left[
     (q-1)(q-2)\frac{{a'}^2 + 1}{a^2}
     - \frac{p(p+q-2)}{q-2}{\phi'}^2 +
    p(p-1)h^2e^{-\frac{2(p+q-2)}{q-2}\phi}
    \right.\\
    &\left. - 2\Lambda e^{-\frac{2p}{q-2}\phi}
    - \frac{e^{\frac{2p(q-1)}{q-2}\phi}}{a^{2(q-1)}}{\psi'}^2
    \right]a^{q-1},
   \end{aligned}
  \end{equation}
  where the boundary terms have vanished since the boundary conditions
  require $a(r_\pm) = 0$, $|a'(r_\pm)| = 1$ and $\phi'(r_\pm) = 0$ 
  at the boundaries $r=r_\pm$.
  Using the equations of motion, we evaluate the Euclidean action
  $I_\mathrm{Euclid}$ on shell and it turns out that 
  \begin{equation}
   I_\mathrm{Euclid} = -S,
  \end{equation}
  as explained in appendix~\ref{app:EuclideanAction}.  
  Also, the equation of motion for the form field is given by 
  \begin{equation}
   \left[\frac{e^{\frac{2p(q-1)}{q-2}\phi}}{a^{q-1}}\psi'\right]' = 0,
  \end{equation}
  and it can be easily integrated as 
  \begin{equation}
   \frac{e^{\frac{2p(q-1)}{q-2}\phi}}{a^{q-1}}\psi' = b,
    \label{eq:definition_b}
  \end{equation}
  where $b$ is an integration constant.
  The total flux $\Phi$ is rewritten as
  \begin{equation}
   \Phi = \Omega_{q-1}\left[\psi(r_+) - \psi(r_-)\right]
    = b \Omega_{q-1}\int^{r_+}_{r_-} \mathrm dr \,
    e^{-\frac{2p(q-1)}{q-2}\phi} a^{q-1}.
    \label{eq:total_flux}
  \end{equation}

We are now ready to derive the first law of de Sitter thermodynamics in
  our setup. 
  We consider the first variation of the action 
  $I_\mathrm{Euclid}[a,\phi,\psi]$ with respect to $a$, $\phi$ and $\psi$.
  Suppose both $\{a,\phi,\psi\}$ and 
  $\{a + \delta a,\phi + \delta \phi, \psi + \delta \psi\}$ are different sets
  of solutions satisfying the equations of motion, the first variation
  of the action $\delta I_\mathrm{Euclid}$ is given by 
  \begin{equation}
   \delta I_\mathrm{Euclid} =
     \left.\frac{\Omega_p\Omega_{q-1}}{8\pi h^p}
       \frac{e^{\frac{2p(q-1)}{q-2}\phi}}{a^{q-1}}\psi' \delta \psi
      \right|^{r_+}_{r_-}
    + \int^{r_+}_{r_-}\mathrm dr
    \left(\text{EOM for $a$, $\phi$ and $\psi$}\right).
  \end{equation}
  The integrand in the last term will vanish because of the
  equations of motion for $a$, $\phi$ and $\psi$. As a result, only the
  boundary term contributes to the first variation of the action.
  By using (\ref{eq:definition_b}) and (\ref{eq:total_flux}) the first
  law can be derived as 
  \begin{equation}
   \mathrm d S = - \frac{\Omega_{p-2}b}{4(p-1)h^p}\mathrm d\Phi.
    \label{eq:first_iaw}
  \end{equation}
  This implies that the entropy is described by a function of the total
  flux.
  Hence the entropy $S$ is a thermodynamic potential with respect to the
  total flux $\Phi$ as a natural thermodynamic variable.
  We shall call a sequence of solutions which satisfies the first law
  a ``branch'' of solutions. 
  
So far,
we have assumed that the bulk cosmological constant $\Lambda$ is
  not a dynamical variable but a given constant. 
  However it is probable that $\Lambda$ is a dynamical variable induced
  by dynamical fields such as a scalar field. 
  In this case we consider $\Lambda$ as an additional thermodynamic
  variable so that the de Sitter entropy is now a function of $\Phi$ and
  $\Lambda$, $S(\Phi,\Lambda)$. The first law (\ref{eq:first_iaw}) is
  easily generalized to include variation of $\Lambda$ as follows. 
  First, dimensional analysis leads to the following scaling relation 
  \begin{equation}
   S(\lambda^{-(q-1)/2}\Phi,\lambda\Lambda)
    = \lambda^{-(p+q-2)/2}S(\Phi,\Lambda).
  \end{equation}
Second, taking derivative 
with respect to $\lambda$ and setting $\lambda =1$, we
  obtain 
  \begin{equation}
   -\frac{q-1}{2}\Phi \frac{\partial S}{\partial \Phi}
    + \Lambda\frac{\partial S}{\partial \Lambda}
    = - \frac{p+q-2}{2}S. 
  \end{equation}
This is the first law of de Sitter thermodynamics for $S(\Phi,\Lambda)$:
  \begin{equation}
   \mathrm dS = - \frac{\Omega_{p-2}b}{4(p-1)h^p}\mathrm d\Phi
    - \left[\frac{p+q-2}{2}S +
       \frac{q-1}{2}\frac{\Omega_{p-2}b}{4(p-1)h^p}\Phi
      \right]\frac{\mathrm d\Lambda}{\Lambda}.
  \end{equation}

  \subsection{Stability}

  In this subsection we discuss thermodynamic stability of the FR branch
  and the warped branch. As we have seen, the entropy $S$ is the
  thermodynamic potential when we choose the total flux $\Phi$ as a 
  natural variable. Therefore, the second law of thermodynamics states
  that, for a fixed value of the total flux, a configuration with larger
  entropy is thermodynamically favored.
  
\begin{figure}[t]
   \begin{center}
    \includegraphics[width=.48\linewidth]{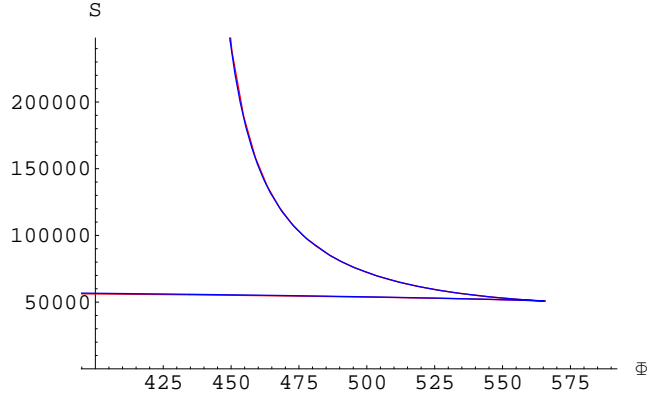}
    \caption{The entropy of the FR branch and the warped branch as
    functions of the total flux $\Phi$.
Difference between entropies of the two branches is so small that two
    lines are indistinguishable in this figure. See
    Fig.~\ref{fig:difference} for the difference. }
    \label{fig:phase}
   \end{center}
  \end{figure}

  To begin with, let us examine thermodynamic property of the FR
  branch. We shall see that the FR branch has two sub-branches and one
  of them is thermodynamically preferred than the other. In the FR
  branch, the entropy $S$ and the total flux $\Phi$ are given by 
  \begin{equation}
   S = \frac{\Omega_{p-2}\Omega_{q}\rho^q}{4h^{p-2}},\quad 
    \Phi = b\Omega_{q}\rho^q
  \end{equation}
  where we have used Eqs.~(\ref{eq:entropy}) and (\ref{eq:total_flux}) with 
  $a(r) = \rho\sin\frac{r}{\rho}$ and $\phi(r) = 0$.
  Note that we can explicitly check that these quantities satisfy the
  first law (\ref{eq:first_iaw}).
  We find that the entropy is written as a function of the total flux
  and splits the FR branch into two sub-branches: a lower-entropy
  sub-branch and a higher-entropy sub-branch, as shown in
  Fig~\ref{fig:phase}. 
  (For example, see \cite{Kinoshita:2007ci}.) 
  Therefore, for a given total flux the 
  higher-entropy sub-branch is preferred than the lower-entropy
  sub-branch within the FR branch. 
  
  The critical point dividing the FR branch into two sub-branches is
  determined as follows. As seen before, the FR branch satisfies the
  first law. However, the entropy $S(\Phi)$ is a double-valued function
  of the total flux $\Phi$. Nonetheless, the FR solutions can be
  described as an one-parameter family of solutions, for example, in
  terms of the Hubble expansion rate $h$ of the external de Sitter
  space. Actually, the entropy is a single-valued function of $h$. These
  facts mean that a map from $\Phi$ to $h$ becomes singular at the
  critical value. Hence we can obtain the critical point by solving 
  $\mathrm d\Phi/\mathrm dh = \mathrm dS/\mathrm dh = 0$, 
  which yields the critical value 
  \begin{equation}
   h_{\mathrm{c}(l=0)}^2 = \frac{2\Lambda(p-2)}{(p-1)^2(p+q-2)},\quad
    b_{\mathrm{c}(l=0)}^2
    = \frac{(p-1)(p+q-2)}{(p-2)(q-1)}h_{\mathrm{c}(l=0)}^2.
  \end{equation}
  By comparing with (\ref{eqn:hcrit0-dynam}), it is easy to see that
  this thermodynamical critical point agrees with the threshold at which
  the $l=0$ mode becomes massless on the FR branch. In addition, the
  lower-entropy sub-branch is dynamically unstable against homogeneous
  ($l=0$) excitation of the internal space. Therefore, we see complete
  agreement between thermodynamic and dynamical stability of the two
  sub-branches. 
  
  In the previous paragraphs we have compared entropies of two
  sub-branches within the FR branch. We now compare entropies of the FR
  branch and the warped branch.  For simplicity we shall consider the
  case with $p=4$ and $q=4$ as an explicit example. 

\begin{figure}[t]
   \begin{center}
    \includegraphics[width=.48\linewidth]{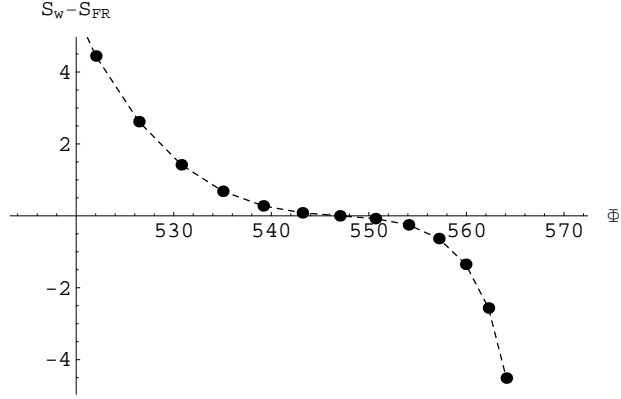}
    \caption{The difference between the entropy of the warped branch
    $S_\mathrm{w}$ and that of the FR branch $S_\mathrm{FR}$.}
    \label{fig:difference}
   \end{center}
  \end{figure}

  We denote the entropy of the FR branch and that of the warped branch
  as $S_\mathrm{FR}$ and $S_\mathrm{w}$, respectively.
  A difference between $S_\mathrm{w}$ and $S_\mathrm{FR}$
  for various values of the total flux is shown in
  Fig.~\ref{fig:difference}. 
  It turns out that for 
  $\Phi < 32\sqrt{3}\pi^2 \Lambda^{-3/2}$, or equivalently, for 
  \begin{equation}
   h^2 < \frac{\Lambda}{18}, \label{eqn:hcrit-thermo}
  \end{equation}
  the warped branch has larger entropy and thus is thermodynamically
  stable. On the other hand, for
  $\Phi > 32\sqrt{3}\pi^2 \Lambda^{-3/2}$, or equivalently, for 
  $h^2 > \frac{\Lambda}{18}$, the FR branch has larger entropy and is
  thermodynamically stable. At the critical point where two branches
  merge the solution is marginally stable. 

  It is worth noting that the thermodynamic stability investigated here
  agrees with the dynamical stability examined in
  Sec~\ref{sec:dynamical_stability}. The FR branch has dynamical 
  instability arising from inhomogeneous ($l=2$) excitation when
  $h^2<h^2_{\mathrm{c} (l=2)}$, where $h^2_{\mathrm{c} (l=2)}$ is given
  by (\ref{eqn:hcrit2-dynam}). On the other hand, the warped branch is 
  dynamically unstable for $h^2>h^2_{\mathrm{c} (l=2)}$. 
  For $p=4$ and $q=4$, the critical value $h^2_{\mathrm{c} (l=2)}$ for
  the dynamical stability agrees with the critical value for the
  thermodynamic stability given in the right hand side of
  (\ref{eqn:hcrit-thermo}). Therefore, we again see complete agreement
  between thermodynamic and dynamical stabilities.

\section{Cosmological implications}
\label{sec:Cosmological_implications}

In previous sections we have investigated stability of two branches of
flux compactification from thermodynamic and dynamical perspectives. One
branch has higher symmetry $\mathrm{SO}(q+1)$, where $q$ is the number of extra
dimensions, and corresponds to unwarped, Freund--Rubin
solutions~\cite{Freund:1980xh}. The other branch has lower symmetry 
$\mathrm{SO}(q)$ and corresponds to warped solutions found recently by
one of the authors~\cite{Kinoshita:2007uk}. By fixing or scaling out the
higher-dimensional cosmological constant, each branch is parameterized
by one parameter, either total flux $\Phi$ of an antisymmetric field or
the Hubble expansion rate $h$ of the $4$-dimensional de Sitter
metric. We have seen that the unwarped branch is dynamically stable for
$h$ larger than a critical value $h_{\mathrm{c}(l=2)}$ but unstable for smaller
values. On the other hand, the warped branch is dynamically unstable for
$h>h_{\mathrm{c}(l=2)}$ and stable for $h<h_{\mathrm{c}(l=2)}$. To investigate thermodynamic
perspective, we have defined the total de Sitter entropy $S$ as the
$4$-dimensional de Sitter entropy integrated over extra dimensions. We 
have shown that the dynamically stable branch, i.e. the unwarped
(or warped) branch for $h>h_{\mathrm{c}(l=2)}$ (or $h<h_{\mathrm{c}(l=2)}$, respectively), always has
larger total de Sitter entropy than the dynamically unstable
branch. Therefore, thermodynamic stability agrees with dynamical
stability.

In this section we consider cosmological implications of these
results. For this purpose we shall first define the $4$-dimensional
Einstein frame. For the ($4+q$)-dimensional metric of the form
%
\begin{equation}
 G_{MN}\mathrm dX^M\mathrm dX^N
  = A^2(x,y)g_{\mu\nu}(x)\mathrm dx^{\mu}\mathrm dx^{\nu}
  + q_{mn}(y)\mathrm dy^m\mathrm dy^n, 
\end{equation}
the higher-dimensional Einstein-Hilbert action is 
%
\begin{equation}
 I_{4+q} = \frac{(M_{4+q})^{2+q}}{2} \int \mathrm d^{4+q}X\sqrt{-G}R[G]
  = \frac{M_4^2}{2}\int \mathrm d^4x\sqrt{-g}\Omega^2 R[g]  + \cdots,
\end{equation}
where $M_{4+q}$ and $M_4$ are ($4+q$)- and $4$-dimensional Planck
scales (in this section we have temporarily restored the Planck scales), and 
%
\begin{equation}
 \Omega^2 = \frac{(M_{4+q})^{2+q}}{M_4^2}\int \mathrm d^qy\sqrt{q}A^2. 
\end{equation}
Since $\Omega$ in general depends on the $4$-dimensional coordinates
$x^{\mu}$, the resulting $4$-dimensional effective theory describing
$g_{\mu\nu}$ is not Einstein but a scalar-tensor theory. It is
convenient to define the $4$-dimensional Einstein frame
$g^{(E)}_{\mu\nu}$ by 
%
\begin{equation}
 g^{(E)}_{\mu\nu} = \Omega^2 g_{\mu\nu},
\end{equation}
in terms of which $I_{4+q}$ now includes the $4$-dimensional
Einstein-Hilbert term:
%
\begin{equation}
 I_{4+q} = \frac{M_4^2}{2}\int \mathrm d^4x\sqrt{-g^{(E)}} R[g^{(E)}]
  + \cdots.
\end{equation}

Suppose that the $4$-dimensional metric $g_{\mu\nu}$ represents a de
Sitter spacetime with the Hubble expansion rate $h$. The corresponding
Einstein frame metric $g^{(E)}_{\mu\nu}$ has the Hubble expansion rate 
%
\begin{equation}
 h_E = \Omega^{-1}h, 
\end{equation}
and the de Sitter entropy
%
\begin{equation}
 S_E = 8\pi^2 M_4^2h_E^{-2}
  = \frac{8\pi^2(M_{4+q})^{2+q}}{h^2}\int \mathrm d^qy\sqrt{q}A^2. 
\end{equation}
Actually, this agrees with the total de Sitter entropy (\ref{eq:entropy}): 
%
\begin{equation}
 S_E = S. 
\end{equation}

For cosmological considerations energy density is more convenient than
de Sitter entropy since the former can easily be extended to a general
FRW universe. The effective energy density in the Einstein frame is
%
\begin{equation}
 \rho_E = 3M_4^2h_E^2 = \frac{3M_4^4 h^2}{(M_{4+q})^{2+q}}
  \left(\int \mathrm d^qy\sqrt{q}A^2\right)^{-1}. 
\end{equation}
This is related to the total de Sitter entropy as
%
\begin{equation}
 \frac{\rho_E}{3M_4^4} = \frac{8\pi^2}{S}. 
\end{equation}

Therefore, the results of previous sections are restated in terms of the
effective energy density $\rho_E$ in the Einstein frame as follows:
dynamically stable branch, i.e. the unwarped (warped) branch for
$h>h_{\mathrm{c}(l=2)}$ ($h<h_{\mathrm{c}(l=2)}$), always has lower
$\rho_E$ than the dynamically unstable 
branch. This strongly suggests that a solution in the dynamically
unstable branch should evolve to a solution in the dynamically stable
branch. The latter solution is uniquely specified by the former since the
flux $\Phi$ conserves. By this evolution, $\rho_E$ decreases.

\begin{figure}[t]
   \begin{center}
    \includegraphics[width=.48\linewidth]{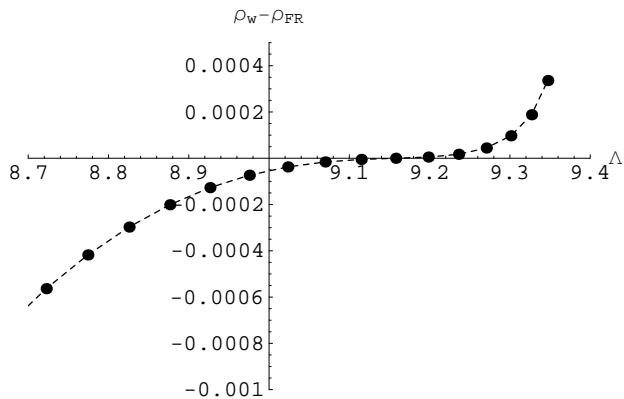}
    \caption{The difference of the effective energy density between the
    warped branch and the FR branch with a fixed total flux.}
    \label{fig:drho_lambda}
   \end{center}
  \end{figure}

Moreover, the results of the previous sections suggest a new type of
phase transition. Suppose that $\Lambda$ is not the genuine constant but
has a contribution from dynamical fields such as a scalar field. In this
case $\Lambda$ is expected to decrease while the flux $\Phi$ stays 
constant. (See Fig.~\ref{fig:drho_lambda}.) 
If we start with $h>h_{\mathrm{c}(l=2)}$ then the unwarped
branch has lower energy density and is stable. Thus a solution in the
unwarped branch should be realized initially. However, as $\Lambda$
decreases, $h$ also decreases and can reach the critical value
$h_{\mathrm{c}(l=2)}$. 
At that point, the stable and unstable branches merge. After that, for
$h<h_{\mathrm{c}(l=2)}$, solutions in the warped branch should be
realized since this branch has lower energy density and is stable. 
This phase transition should be second-order since only one of the two
branches is stable at a given time. 

Second-order phase transitions play important roles in cosmology. For
example, the end of hybrid inflation is due to second-order phase
transition. Thus, this kind of phase transition may provide a new way of
realizing hybrid inflation in higher-dimensional theories. This
possibility will be investigated in future publications.

 \section{Discussion}
 
 In this paper we have investigated stability of two branches of
 Freund--Rubin compactification from two perspectives; one is
 thermodynamic stability based on de Sitter entropy, and the other is
 dynamical stability of linear perturbations around the background
 solutions. 
 
 We have analyzed linear perturbations around the warped solutions in 
 order to examine dynamical stability of the solutions.
 The warped solutions are stable if the Hubble expansion rate of the
 external de Sitter spacetime is low enough. In the same regime of the
 Hubble expansion rate, the Freund--Rubin solutions have instability
 arising from the $l=2$ mode. It follows from what has been said thus
 far that deformation of the internal space and warping will stabilize
 unstable configurations. This may be considered as spontaneous breaking 
 of the symmetry of internal space in the sense that less symmetric
 configurations are dynamically chosen. Actually, for the reason
 explained below, this phenomenon is natural from gravitational
 viewpoints.  In cosmology it is well known that configurations with
 high matter density suffer from gravitational instability due to
 long-wavelength modes, namely Jeans instability. Jeans instability
 develops inhomogeneities in the universe and results in structure 
 formation. In the case of FR compactification, Einstein equation tells
 that the external space with small Hubble expansion rate corresponds to
 the internal space with large flux density. Therefore, it is natural to
 expect that when the Hubble expansion rate is low enough, the flux
 distribution may become inhomogeneous due to analogue of the Jeans
 instability. 

 In order to analyze thermodynamic properties, we have first derived the
 first law of de Sitter thermodynamics in terms of entropy and total
 flux. Sequence of solutions belonging to a branch obey this first law
 when those parameters characterizing the solutions change. Since the
 entropy is a natural thermodynamic potential for the total flux, for a
 given total flux, configuration with higher entropy should be favored 
 thermodynamically. For $p=4$ and $q=4$ we have compared the entropy of
 the FR branch with that of the warped branch. There is a critical value
 of the Hubble expansion rate at which two branches merge and it is thus
 obvious that the entropies of the two branches agree at the critical
 value. For smaller Hubble expansion rate, the entropy of the warped
 branch is larger than that of the FR branch and thus the warped branch
 is thermodynamically favored. On the other hand, for larger Hubble
 expansion rate, the FR branch is thermodynamically favored. We found
 complete agreement of thermodynamic stability and dynamical stability.

 It is intriguing to see that correlation between thermodynamic
 stability and dynamical stability exists for Freund--Rubin flux
 compactification. For a certain class of black objects the existence of
 such correlation has been known, that is the so-called Gubser--Mitra 
 conjecture. This conjecture has been explicitly checked to hold for
 various black strings and branes. From what has been discussed above,
 it is probably natural to expect that the concept of correlated
 stability can be extended to a wider class of gravitating systems.

 \begin{acknowledgments}
  We would like to thank Jiro~Soda, Masaru~Shibata, Takahiro~Tanaka and
  Tetsuya~Shiromizu for valuable comments.
  We are deeply grateful to Katsuhiko Sato for his continuous support.
  The work of SK was in part supported by JSPS through a Grant-in-Aid.
  The work of SM was supported in part by MEXT through a Grant-in-Aid for
  Young Scientists (B) No.~17740134, by JSPS through a Grant-in-Aid for
  Creative Scientific Research No.~19GS0219 and through a Grant-in-Aid for
  Scientific Research (B) No.~19340054, and by the Mitsubishi Foundation.
  This work was supported by World Premier International Research Center
  Initiative (WPI Initiative), MEXT, Japan. 
 \end{acknowledgments}
 
 \appendix
 
 \section{Gauge}
\label{app:gauge} 

In the background spacetime given by 
 \begin{equation}
  g_{MN}\mathrm dx^M\mathrm dx^N
   = A^2(y)g_{\mu\nu}\mathrm dx^\mu\mathrm dx^\nu + \mathrm dy^2 
   + B^2(y)\gamma_{ij}\mathrm dz^i\mathrm dz^j,
 \end{equation}
 where $g_{\mu\nu}$ and $\gamma_{ij}$ are respectively the metric of
 $p$-dimensional de Sitter space and $(q-1)$-dimensional round sphere, 
 we consider scalar-type perturbations,
 \begin{equation}
  \begin{aligned}
   \delta g_{\mu\nu} &= 
   2A^2\left(\nabla_\mu\nabla_\nu - \frac{1}{p}g_{\mu\nu}\nabla^2\right)
   H^{(\mathrm{LL})} + A^2 g_{\mu\nu} H^{(\mathrm Y)},\\
   \delta g_{\mu y} &= \nabla_\mu H_y^{(\mathrm L)},\\
   \delta g_{ij} &= 
   2B^2\left(D_iD_j - \frac{1}{q-1}\gamma_{ij}D^2\right)
   h^{(\mathrm{LL})} + B^2 \gamma_{ij} h^{(\mathrm Y)},\\
   \delta g_{i y} &= D_i h_y^{(\mathrm L)},\\
   \delta g_{yy} &= h_{yy},\\
   \delta g_{\mu i} &= \nabla_\mu D_i \eta.
  \end{aligned}
 \end{equation}
 Note that $\nabla_\mu$ and $D_i$ denote covariant derivatives
 associated with $g_{\mu\nu}$ and $\gamma_{ij}$, respectively.
 
 These components transform as 
 \begin{equation}
  \begin{aligned}
   H^{(\mathrm{LL})} &\to H^{(\mathrm{LL})} - A^{-2}\xi^{(\mathrm{L})}, &
   H^{(\mathrm{Y})} &\to H^{(\mathrm{Y})}
   - A^{-2}\frac{2}{p}\nabla^2\xi^{(\mathrm{L})} - 2\frac{A'}{A}\xi_y, &
   H^{(\mathrm{L})} &\to H^{(\mathrm{L})}
   - \xi_y - A^2(A^{-2}\xi^{(\mathrm{L})})',\\
   h^{(\mathrm{LL})} &\to h^{(\mathrm{LL})} - B^{-2}\xi^{(\mathrm{l})},
   &
   h^{(\mathrm{Y})} &\to h^{(\mathrm{Y})}
   - B^{-2}\frac{2}{q-1}D^2\xi^{(\mathrm{l})} - 2\frac{B'}{B}\xi_y,&
   h^{(\mathrm{L})} &\to h^{(\mathrm{L})}
   - \xi_y - B^2(B^{-2}\xi^{(\mathrm{l})})',\\
   h_{yy} &\to h_{yy}
   - 2{\xi_y}',&
   \eta &\to \eta - \xi^{(\mathrm{L})} - \xi^{(\mathrm{l})},
  \end{aligned}
 \end{equation}
 
 By setting 
 \begin{equation}
  \xi^{(\mathrm{L})} = A^2 H^{(\mathrm{LL})}, \quad
   \xi_y = H^{(\mathrm{L})} - A^2 (H^{(\mathrm{LL})})',
 \end{equation}
 and assuming that the perturbations does not depend on
 $z^i$-coordinates, we can simplify the form of metric perturbations so
 that non-vanishing components are 
 \begin{equation}
  \delta g_{\mu\nu} = A^2 g_{\mu\nu} H^{(\mathrm Y)},\quad
   \delta g_{ij} = B^2 \gamma_{ij} h^{(\mathrm Y)},\quad
   \delta g_{yy} = h_{yy},
 \end{equation}
 and the other components vanish.  
 In addition, using parts of the linearized Einstein equations for
 scalar-type perturbations we have an algebraic relation 
 \begin{equation}
  (p-2)H^{(\mathrm Y)} + h_{yy} + (q-1)h^{(\mathrm Y)} = 0.
 \end{equation}
 
 Finally, we expand the perturbations in harmonics $\mathsf Y(x)$ on the
 $p$-dimensional de Sitter space $g_{\mu\nu}$, and then we define two
 variables $\Pi(y)$ and $\Omega(y)$ as  
 \begin{equation}
  H^{(\mathrm Y)} = \Pi\mathsf Y(x), \quad 
   h_{yy} = (\Pi - \Omega)\mathsf Y(x), \quad
   h^{(\mathrm Y)} = \left(\frac{\Omega}{q-1}-\frac{p-1}{q-1}\Pi\right)
       \mathsf Y(x).
 \end{equation} 
 
 \section{Euclidean action and entropy}
 \label{app:EuclideanAction}

 The Euclidean action is given by 
   \begin{equation}
   \begin{aligned}
    I_\mathrm{Euclid}[a,\phi,\psi]
    =& - \frac{\Omega_p\Omega_{q-1}}{16\pi h^p}
    \int \mathrm dr
    \left[
     (q-1)(q-2)\frac{{a'}^2 + 1}{a^2}
     - \frac{p(p+q-2)}{q-2}{\phi'}^2 +
    p(p-1)h^2e^{-\frac{2(p+q-2)}{q-2}\phi}
    \right.\\
    &\left. - 2\Lambda e^{-\frac{2p}{q-2}\phi}
    - \frac{e^{\frac{2p(q-1)}{q-2}\phi}}{a^{2(q-1)}}{\psi'}^2
    \right]a^{q-1}\\
    =& - \frac{\Omega_p\Omega_{q-1}}{16\pi h^p}
    \int \mathrm dr \mathcal L(a,a',\phi,\phi',\psi').
   \end{aligned}
  \end{equation}
  Here, since this action is invariant under the following transformation:
  \begin{equation}
   h \to \lambda h, \quad \phi \to \phi + \ln \lambda, \quad 
    a \to \lambda^{p/(q-2)} a, \quad r \to \lambda^{p/(q-2)} r, \quad 
    \psi \to \psi,
  \end{equation}
  where $\lambda$ is an arbitrary constant, we have an identity
  \begin{equation}
   \left[
    \frac{p}{q-2}\frac{\partial \mathcal L}{\partial a'}a + 
    \frac{\partial \mathcal L}{\partial \phi'}
   \right]'
   - p \mathcal L
   + 2p(p-1)h^2e^{-\frac{2(p+q-2)}{q-2}\phi}a^{q-1} = 0,
  \end{equation}
  provided that $a(r)$, $\phi(r)$ and $\psi(r)$ satisfy the equations of
  motion. Hence the on-shell Euclidean action can be written as 
  \begin{equation}
   \begin{aligned}
   I_\mathrm{Euclid} =& - \frac{\Omega_p\Omega_{q-1}}{16\pi h^p}
    \int \mathrm dr \mathcal L(a,a',\phi,\phi',\psi')\\
    =& - \frac{\Omega_p\Omega_{q-1}}{8\pi h^{p-2}}
    \int \mathrm dr (p-1)e^{-\frac{2(p+q-2)}{q-2}\phi}a^{q-1}
    \equiv - S,
   \end{aligned}
  \end{equation}
  where the boundary terms vanish because of the boundary conditions. 
  It is easy to show by using $(p-1)\Omega_p = 2\pi\Omega_{p-2}$ that
  $S$ agrees with the de Sitter entropy defined in (\ref{eq:entropy}).

\end{document}